\documentstyle[a4,12pt]{article}
  \newfont{\sect}{cmbx12}
  \newcounter{Intro}
  \newcounter{Partie}
  \newcommand{\nsection}[1]{\begin{center}
				{\sect{\Roman{Partie}. }}{{\sect #1}}
				\stepcounter{Partie}
			    \end{center}}
  \newcommand{\nappendix}[1]{{\appendix #1}}
  \newcounter{sPartie}
  \newcommand{\nssection}[1]{\begin{flushleft}
				\addtocounter{Partie}{-1}
				{\sect{\Roman{Partie}.\thesPartie }}
							  {{\sect #1}}
				\stepcounter{sPartie}
			    \end{flushleft}}

\begin{document}

\begin{titlepage}
   \topmargin 2cm
   \begin{trivlist}
      \item[] \centerline{\Large{\bf{Singular Hypersurfaces in}}}
      \item[] \centerline{\Large{\bf{Scalar-Tensor Theories of Gravity}}}
      \itemsep 0.5cm
      \item[] \centerline{\normalsize{C. Barrab\`es$^{\dagger\ddagger}$
		\footnote{barrabes@celfi.phys.univ-tours.fr}
                and G.F. Bressange$^{\dagger}$
		\footnote{bressang@celfi.phys.univ-tours.fr}}}
      \itemsep 0cm
      \item[] 	\begin{center}
			$^{\dagger}$UPRES A 6083 du CNRS\\
      			Facult\'e des Sciences et Techniques\\
      			Parc de Grandmont 37200 Tours - France
		\end{center}
      \itemsep 1cm
      \item[] \begin{center}
	 ${\ddagger}$ D\'epartment d'Astrophysique Relativiste et Cosmologie,\\
           UPR 176 du CNRS, Observatoire de Paris,\\
           92190 Meudon, France
	    \end{center}
	\end{trivlist}
      \vspace{2cm}
\newpage
        \begin{center}
	\begin{trivlist}
        \headsep 13cm
      \item[] \centerline{\bf{Abstract}}
      \itemsep 0cm
      \item[]
       \hspace{0.7cm}
	   We study singular hypersurfaces in tensor multi-scalar
       theories of gravity. 
       We derive in a distributional and then in an intrinsic way, the general
       equations of junction valid for all types of hypersurfaces, 
       in particular for lightlike shells and
       write the general equations of evolution for these objects.
       We apply this formalism to various examples
       in static spherically symmetric spacetimes,
       and to the study of planar domain walls and plane impulsive waves.
  \end{trivlist}
  \end{center}
\end{titlepage}
\newpage
\nsection{INTRODUCTION}

   Most of the various attempts to quantize the gravitational
field have led to the conclusion that at Planckian energies the
Einstein theory of gravity has to be extended in order to include
scalar fields. In the low energy limit, string theory gives back
classical general relativity with a scalar field partner (the dilaton)
and the effective action shows that the dilaton couples to the
scalar curvature and to the other matter fields \cite{FTCa}. Scalar fields 
(compactons) also arise in the process of dimensional reduction
of Kaluza-Klein theories \cite{VCho}, and it has been shown that the
presence in the action of high-order terms in the curvature
and its derivative amounts to introducing scalar fields with
appropriate potentials in the Einstein-Hilbert action of general
relativity \cite{TTWGS}.

   The importance of the role that scalar fields could play in
a full theory of gravity has been noticed since a long time \cite{Jo}.
The pioneering works of Fierz, Jordan, and Brans and Dicke \cite{FJBD} have opened
on the first scalar-tensor theory of gravity (usually referred to
as the the Brans-Dicke theory) which includes besides the gravitational
field $g_{\mu \nu}$, a massless scalar field $\varphi$ and a free
parameter $\omega$. This theory was later generalized \cite{BNW} by making
the parameter field-dependent, i.e. $\omega(\varphi)$, and by introducing
a potential term $V(\varphi)$. More recently, using nonlinear $\sigma$-models
multi-scalar-tensor theories have been considered and their predictions
have been dicussed and compared with general relativity in the
weak-field and strong-field regimes \cite{DEFB}. All these alternative theories
of gravity belong to the class of scalar-tensor theories in the sense
that all the other fields (generically denoted by ${\Psi}_m$) 
exhibit a universal metric coupling to the gravitational field with
the same metric tensor. On the other hand the scalar fields (dilatons,
compactons) which appear in the string and Kaluza-Klein theories
have a non-metric coupling with the fields ${\Psi}_m$, and instead
induce a local spacetime dependence of the coupling constants. This entails
fundamental differences with general relativity, in particular with the
equivalence principle \cite{Da}.

   Besides these theoretical considerations on the possible role played by
scalar fields in a complete theory of gravity, an important motivation for 
considering scalar fields is their application in cosmology to the
scenarios of inflation and the formation of topological defects (monopoles,
cosmic strings and domain walls). A basic
ingredient of all these studies is the introduction of one (or several)
scalar fields (inflatons) which trigger the production of 
phase transitions in the early history
of the universe. Many different models of inflation have been proposed \cite{LVOM},
some of them within the framework of scalar-tensor or string theories
of gravity. For instance, in extended inflation \cite{LaSt}, general relativity is
replaced by the Brans-Dicke theory,  and in a later version sometimes referred 
to as hyperextended inflation \cite{StABM} the parameter $\omega$ 
varies with the scalar field $\varphi$. The properties of
scalar-tensor cosmological models have been much studied: methods for 
obtaining exact solutions of the field equations have been given \cite{BWK},
constraints for succesful extended inflation and constraints from inflation
on scalar-tensor theories have been formulated \cite{W}, and the existence of
an attractor mechanism towards general relativity have been discussed \cite{DN}.  
The formation and dynamics of spherical bubbles of true vacuum in the 
Brans Dicke theory have been studied in a thin-wall
formalism \cite{GZ},\cite{SM}. String cosmologies have also aroused much
interest \cite{V} and solutions to the peculiar difficulties
associated with the dilaton have been proposed \cite{PDV}. A class
of supersymmetric domain walls in $N=1$ supergravity and within
effective string theories have been obtained and their gravitational
effects have been described \cite{Cv}. 

    In this paper we study the junction conditions which have to be
satisfied by the various fields at an arbitrary singular hypersurface
separating two different spacetimes in scalar-tensor 
theories of gravity. This work generalizes previous
descriptions of thin shells in the Brans Dicke theory 
\cite{GZ},\cite{SM},\cite{Su},\cite {SW}, and \cite{LW} in the
sense that we present a general algorithm wich can 
be applied to a singular hypersurface of any type (timelike,
spacelike or lightlike) and we consider the possibilty of having discontinuous
gauge fields. The results obtained in the
timelike case apply to arbitrary surface layers and in
particular to domain walls as considered in the refs \cite{GZ}-\cite{Cv}, with
eventually surface currents as it may be the case for some superconducting 
domain walls coming from a supersymmetric action \cite{Mo}.
The less considered spacelike case might
for instance correspond to a transition layer which suddenly appears
and disappears all over space at a given time -examples of this situattion
can be found in \cite{FMML}. The lightlike case
has interesting properties because it can at the same time describe a
lightlike shell with surface energy density and surface stresses, and an
impulsive gravitational wave which is accompanied by
shock waves when discontinuous gauge fields are present. These
waves have been shown to be of some interest in string theory \cite{Gu} as 
plane waves are exact classical solutions at all order of the
string tension parameter \cite{GuHS}.
 
   It is known that in scalar-tensor and dilatonic theories two 
conformally related metrics can be used, the Jordan-Fierz or string
metric and the Einstein metric. The Jordan-Fierz or string metric
is usually referred to as the physical metric as the
stress-energy tensor for the matter fields is conserved
in this metric and not in the other one -see however Cho in ref.\cite{VCho}.
However many of the mathematical properties of these
theories (asymptotic behavior, Cauchy problem...) are
more conveniently investigated in the Einstein metric. 
While most of the authors having studied thin shells in the 
Brans-Dicke theory have worked in the Jordan-Fierz frame
we have prefered here to use the Einstein frame. The main reasons for making
such a choice is that it offers a
simpler set of equations, and it enables an easier comparison with previous results
obtained in general relativity \cite{BI},\cite{BBH}.

   This paper is organized as follows. In section 2 we give a brief
survey of the scalar-tensor theories of gravity and in section 3
we present our general results concerning the junction conditions
accross an arbitrary hypersurface. These conditions are described
in a distributional formalism and within an intrinsic approach,
and the existence of discontinuous gauge fields is considered.
In the next two
sections we consider various examples illustrating the general formalism 
presented in section 3. These applications
concern spherically symmetric shells (section 4), and planar
shells and plane impulsive waves (section 5). 
In the last section we briefly discuss the differences which appear in
the description of a shell when using dilatonic theories instead
of scalar-tensor theories of gravity. Finally some static
spherically symmetric solutions of the Brans-Dicke theory are
presented in appendix A.

{\it Conventions:} Our metric signature is $( - + + +)$ and we use
the standard conventions for the Riemann tensor of Misner, Thorne
and Wheeler \cite {MTW}. Greek indices run from $0$ to $3$.
\vspace{0.5cm}
\nsection{SURVEY OF SCALAR-TENSOR THEORIES OF GRAVITATION}

	The scalar-tensor theories of gravitation
are alternative theories of gravity
which generalize in the most natural way the
Brans-Dicke theory by introducing a finite number
of scalar fields,
${\varphi}^i, i = 1,2,..n $, each characterized by
a particular coupling constant to local matter
-see for instance Damour and Esposito-Far\`{e}se \cite{DEFB} 
for a review of scalar-tensor theories.
These theories are covariant tensor field theories 
and they coincide with general relativity in the
post-newtonian approximation. They are metrics theories,
which means
that the matter fields are minimally coupled to a universal
covariant 2-tensor,
$\,\bar{g}_{\mu \nu}\,$, usually referred to as the physical 
metric or the Jordan-Fierz metric.
They can also be described within another frame, which
is conformally related to the previous one, in such a way
that the Einstein-Hilbert
term is recovered in the action. This frame is called the Einstein
conformal-frame and the corresponding metric $\,g_{\mu \nu}\,$, 
the Einstein metric. The relation between the two metrics is 
	\begin{equation}
	\bar{g}_{\mu \nu}\,= \,A^{2}({\varphi}) g_{\mu \nu}\hspace{1cm},
  	\end{equation}
where the conformal factor $A^{2}({\varphi})$ is a smooth function 
of the n scalar fields $\,{\varphi}^{i}\,$.

  The general form of the action
for scalar-tensor theories of gravity is, in the Einstein conformal-frame
\begin{equation}
    S=\frac{1}{16\pi G} \int {d^4 x \sqrt{-g}\; \lbrack R
    -2g^{\mu \nu}{{\partial}_{\mu}}{{\varphi}^{i}}{{\partial}_{\nu}}
    {{\varphi}^{j}}{\gamma}_{ij}-4B({\varphi}) \rbrack}
    +S_{m}\lbrack {\psi}_{m},A^{2}({\varphi})g_{\mu \nu} \rbrack .
 \end{equation}
The second term $S_{m}$ is the action for the ``matter'' fields
(fermionic or bosonic), 
collectively denoted by $\,{\psi}_{m}\,$ which as indicated in $S_m$
couple to the
Jordan-Fierz metric. In the first term, the scalar fields $\varphi^i$ 
appear in a non-linear $\sigma$-model. This means that the
quantities $\varphi ^i$ can be 
viewved as internal local coordinates in a n-dimensional manifold (
target space )
endowed with a metric $d\sigma ^2$ which is written
	$$d \sigma ^{2}=
	\gamma ^{ij}(\varphi)d \varphi _{i}d \varphi _{j}\hspace{1cm}.$$
in the local coordinate system $\lbrace \varphi^i\rbrace$.

	The potential term $B$ is a smooth function (at least $\,C^{2}\,$)
of the scalar-fields $\,{\varphi}^{i}\,$ and  may include a cosmological
term.
The field equations for the metric tensor $g_{\mu \nu}$ and the scalar
fields ${\varphi}^{i}$ which follow from the above action are respectively
  \begin{equation}
  G_{\mu \nu}=8\pi G \left( T_{\mu \nu}+T_{\mu \nu}(\varphi)\right)\,
		\equiv \, 8\pi G {\cal T}_{\mu \nu}\hspace{1cm},
  \end{equation}
and
 \begin{equation}
     \Box{\varphi}^{i}+{\gamma}_{jk}^{i} \: g^{\mu \nu} {\partial}_{\mu}
     {\varphi}^{j}{\partial}_{\nu}{\varphi}^{k}
     -{\beta}^{i} ({\varphi})
	      =  -4 \pi G {\alpha}^{i}({\varphi})\: T\hspace{1cm},
  \end{equation}
where $\Box$ is the d'Alembertian with respect to the metric $g$.
( Note that eq.(4) is not covariant with respect to the indices of the
target space ). ${\cal T}_{\mu \nu}$ is the total stress-energy tensor, $T_{\mu \nu}$ is
the stress-energy tensor of the matter fields (see below) and  
$T = T^{\mu}_{\mu}$ its trace, and 
\begin{equation}
 T_{\mu \nu}(\varphi)\,\equiv\,\frac{{\gamma}_{ij}}{4\pi G}\,\lbrack \,{\partial}_{\mu}\: {\varphi}^{i}
				    \: {\partial}_{\nu}\: {\varphi}^{j}
        -\frac{1}{2} \,g_{\mu \nu}\ ( g^{\alpha \beta} {\partial}_{\alpha}\: {\varphi}^{i}
					  \: {\partial}_{\beta}\: {\varphi}^{j}) \,\rbrack
	-\frac{B({\varphi})}{4\pi G} \, g_{\mu \nu}
  \end{equation}  
is the stress-energy tensor of the scalar-fields.
The ${\gamma}_{jk}^{i} \: $'s are the Christoffel symbols associated with
the $\sigma$-model metric ${\gamma}_{ij} \: $ and the scalar-fields indices
are raised or lowered with this metric or its inverse.
An other convenient form of (3) is
\begin{equation}
R_{\mu \nu}=8\pi G \left(T_{\mu \nu}-\frac{T}{2}g_{\mu \nu} \right)
		+{\gamma}_{ij}{\partial}_{\mu}
     {\varphi}^{i}{\partial}_{\nu}{\varphi}^{j}\hspace{1cm},     
\end{equation}
We have introduced in (4) the following space-time scalars
\begin{equation}
     {\beta}_{i} (\varphi)=
	     \frac{\partial B (\varphi)}{\partial {\varphi}^{i}} \hspace{1cm} , 
     \hspace{1cm}{\alpha}_{i} (\varphi)=
	     \frac{\partial \ln A (\varphi)}{\partial {\varphi}^{i}}\hspace{1cm},
  \end{equation}
and it is evident from (4) that the $\,{\alpha}_{i} (\varphi)\,$'s 
represent coupling factors
of the scalar fields to matter.

        The stress-energy tensor 
of the matter fields is defined by
  \begin{equation}
     T^{\mu \nu}\:\equiv \frac{2}{\sqrt{-g}}\: \frac{\delta \:
				       S_{m}}{\delta \: g_{\mu \nu}}\hspace{1cm}.
  \end{equation}
In the Einstein frame it is not conserved 
but instead satisfies 
  \begin{equation}
     {\nabla}_{\nu} T^{\mu \nu} =
		{\alpha}_{i} (\varphi) T {\nabla}^{\mu} {\varphi}^{i}\hspace{1cm}. 
  \end{equation}

	Although this work is performed in the Einstein frame, let us  briefly recall 
for the sake of completeness some properties of the Jordan-Fierz description.	
In the Jordan-Fierz frame, only one scalar field $\,{\bar \varphi}\,$ 
can be considered and the action takes the general form 
\begin{equation}
     S=\frac{1}{16\pi G} \int {d^4 x \sqrt{-\bar{g}}\: \lbrack
     \: {\bar \varphi} \bar{R} -\frac{\omega ({\bar \varphi})}{{\bar \varphi}}\:
     {\bar{g}}^{\mu \nu} \: {\partial}_{\mu} {{\bar \varphi}}\:
     {\partial}_{\nu} {{\bar \varphi}}+2{\bar \varphi} \: \Lambda ({\bar \varphi})
	 \: \rbrack}
     +S_{m}\lbrack \: {\psi}_{m},{\bar{g}}^{\mu \nu}\:  \rbrack
  \end{equation}
The field equations for $\:{\bar g}_{\mu \nu}\:$ and $\:{\bar \varphi}\:$
are respectively
   \begin{eqnarray}
      {\bar R}_{\mu \nu}-\frac{\bar R}{2}\:{\bar g}_{\mu \nu}\:
      & = & \frac{8\pi G}{{\bar \varphi}}\: {\bar T}_{\mu \nu}
		  +\Lambda ({\bar \varphi})\:{\bar g}_{\mu \nu}  \: \nonumber\\
      & + & \: \frac{\omega ({\bar \varphi})}{{{\bar \varphi}}^{2}}\: 
	\left[ {\partial}_{\mu}
       {{\bar \varphi}}\: {\partial}_{\nu} {{\bar \varphi}}-\frac{1}{2}
	\: {\bar g}_{\mu \nu}
       \: ({\bar g}^{\alpha \beta}{\partial}_{\alpha} {{\bar \varphi}}\:
       {\partial}_{\beta} {\bar \varphi}) \right] \nonumber\\
      & + & \frac{1}{{\bar \varphi}}\: ({\bar{{\nabla}}_{\mu} 
	\bar{{\nabla}}_{\nu}{\bar \varphi}}
	-{{\bar g}_{\mu \nu}}\: {\Box}^{\bar{}}\: {{\bar \varphi}})\hspace{1cm},
   \end{eqnarray}
and
   \begin{equation}
      {\Box}^{\bar{}}\: {{\bar \varphi}}+\frac{1}{2} {\bar g}^{\alpha \beta}\:
      {\partial}_{\alpha}{{\bar \varphi}}\: {\partial}_{\beta} {{\bar \varphi}}
      \frac{d}{d{\bar \varphi}}\ln (\frac{\omega ({\bar \varphi})}{{\bar \varphi}})
      +\frac{{\bar \varphi}}{\omega ({\bar \varphi})}\: \left[ \frac{\bar R}{2}\:
		+\: \frac{d}{d{\bar \varphi}} ({\bar \varphi} 
	\Lambda ({\bar \varphi}))\right] \: =0 \hspace{1cm},
   \end{equation}
where $\: {\bar T}_{\mu\nu}\:$ is the stress-energy tensor of the matter
fields 
   \begin{equation}
    {\bar T}^{\mu \nu}\: \equiv  \frac{2}{\sqrt{-\bar g}}\: \frac{\delta \:
				       S_{m}}{\delta \:{\bar g}_{\mu \nu}}\hspace{1cm},
   \end{equation}
and where $\,\bar{\nabla}\,$ and $\,{\Box}^{\bar{}}\,$ stand respectively
for the covariant derivative and the d'Alembertian associated
with the metric $\,{\bar g}_{\mu \nu}\,$. As the matter fields couple directly
to the Jordan-Fierz metric their stress-energy tensor is conserved,
i.e. ${\bar{\nabla}}_{\nu} {\bar T}^{\mu \nu}\: =\: 0 $.

	We have the following relations between the two descriptions
  \begin{equation}
    \left\lbrace
      \begin{array}{c}
	 {{\bar \varphi}}^{-1} \ \ \ = \: \: G\  \   A^{2} (\varphi)\\
	 2\: \omega ({\bar \varphi})+3 =  \  \  {\alpha}^{-2} (\varphi)\ \ \ \ \ \\
	 2\: \Lambda ({\bar \varphi}) \ \  =  - B(\varphi)\: A^{-2} (\varphi)
      \end{array}
    \right.
   \end{equation}
and between the matter stress-energy tensors (8) and (13)
\begin{equation}
    T^{\mu \nu} = A^{6}({\varphi}) {\bar T}^{\mu \nu} \hspace{1cm}.
\end{equation}
The Brans-Dicke theory corresponds to the particular case where
$\,\omega ({\bar \varphi})\:= const.\,$ and $\,\Lambda\,=\,0\,$ in the
Jordan-Fierz description, or $\,\alpha\,=\,const.\,$ and $\,B\,=\,0\,$
in the Einstein description. In this case, one obtains from (7)
\begin{equation}
A(\varphi)\: =\:  e^{\alpha \varphi} \hspace{1cm}.
 \end{equation}

\vspace{1cm}
\nsection{SINGULAR HYPERSURFACES IN SCALAR-THEORIES OF GRAVITATION}

	A thin shell corresponds to the situation where there exists a surface 
in the vicinity of which the distribution of stress and energy is so 
strongly concentrated that the thin limit approximation can be used. In
spacetime this yields a three-dimensional hypersurface (timelike,
spacelike or lightlike) along which the metric tensor is only $C^{0}$, 
but ${C^3}$ elsewhere.
The description of thin shells in general relativity and in the timelike and
spacelike cases is well known since the pionneering works of 
Lanczos \cite{Lan} and Israel \cite{Is} and an extension to the
null or lightlike case has recently been done \cite{BI}. It is the purpose of this
section to present a general algorithm, similar to the one
described in \cite{BI}, and adapted to scalar-tensor theories of gravitation
(the case of theories with a dilaton will be studied in section 4).

	Let $\,\Sigma\,$ be the singular hypersurface in space-time corresponding
to the thin shell. As the metric is only $C^0$ on $\Sigma$ 
the total stress-energy tensor
\begin{equation}
   {\cal T}_{\mu \nu}\: =\: T_{\mu \nu}+\,\frac{{\gamma}_{ij}}{4\pi G}(  \: {\partial}_{\mu}\: {\varphi}^{i}
					\: {\partial}_{\nu}\: {\varphi}^{j}
	-\frac{g_{\mu \nu}}{2} \:g^{\alpha \beta}
				   \: {\partial}_{\alpha}\: {\varphi}^{i}
				   \: {\partial}_{\beta}\: {\varphi}^{j}  )
	-\frac{B(\:{\varphi})}{4\pi G}\: g_{\mu \nu}
\end{equation}
which appears in the right-hand side of (3) 
necesseraly contains a $\,\delta\,$-term with support 
on $\,\Sigma\,$. Comparing the eqs (17) and (4) one easily sees that
the singular $\delta$-term can only come from the matter part $T_{\mu \nu}$
of the total stress-energy tensor ,and that because of the presence of the
trace $T$ in the r.h.s. of (4) the scalar fields $\varphi^{i}$ are
$C^0$ on $\Sigma$ and $C^3$ elsewhere. Therefore the metric and the
scalar fields have the same smoothness properties all over spacetime,
as expected from the metric coupling of the scalar fields to
gravity. For instance,
in the scenario of extended inflation, there exists besides the Brans-Dicke 
scalar field another scalar field
(the inflaton) which is responsible for the formation
of true-vacuum bubbles and yields the $\delta$-term in $T_{\mu \nu}$, 
while the role of the Brans-Dicke 
scalar field is to slow down the expansion of the universe.

	There exists two equivalent ways of describing shells which we now 
present - some of the results being common to those obtained in general
relativity we refer the reader
to \cite{BI} for more details.
The first description is based on a four-dimensional distributional approach
and requires a common set of coordinates covering both sides of the shell.
The second one is purely intrinsic, and  allows an independent
and arbitrary choice of coordinates in the two sides of the shell.
The existence of an impulsive gravitational wave and of discontinuous
gauge fields will also be examined within these two descriptions. 
\\
\nssection{Junction conditions : the distributional description}

	We consider a general smooth hypersurface $\: \Sigma \: $ separating two
spacetimes $\: {\cal M}_{+} \: $ and $\: {\cal M}_{-} \: $, endowed with
the metrics $\: g^{+}\: $ and $\: g^{-}\: $
(at least of class $\: C^{3}\: $) and with the scalar fields
$\,{\varphi}^{i}_{+}\,$ and $\,{\varphi}^{i}_{-}\,$
(at least of class $\: C^{2}\: $)
and we introduce a common system of coordinates
$\,x^{\mu}\,$. The matter stress-energy tensors are 
$\: T^{+}_{\alpha \beta}\: $ and $\: T^{-}_{\alpha \beta}\:$, and 
the fields equations (3) and (4) are satified in each domain.
The hypersurface $\: \Sigma \:$ results 
from an isometric soldering of the boundaries 
$\: {\Sigma}^{+}\:$ and $\:{\Sigma}^{+}\:$ which are respectively imbedded
in $\: {\cal M}_{+} \: $ and $\: {\cal M}_{-}\,$. Denoting by 
$\,\left[\: F \:\right] = F^{+} - F^{-} $
the jump across $\: \Sigma \: $ of an arbitrary discontinuous
function $F$, we thus have
\begin{equation}
	   \left\lbrack\: g_{\alpha \beta}\: \right\rbrack \: =\: 0
	   \hspace{0.5cm},\hspace{0.5cm}
	   \left\lbrack\: {\varphi}^{i}\: \right\rbrack \: =\: 0\hspace{1cm}.
\end{equation}
Let $\,\Phi (x)=0\,$ be the equation of $\:{\Sigma}\:$
where $\,\Phi\,$ is a smooth function (at least $\,C^1\,$) 
taking positive (resp. negative) 
values in $\: {\cal M}_{+} \: $ (resp. $\: {\cal M}_{-} \: $) and 
let $\,n\,$ be the normal vector to $\: \Sigma \: $ 
pointing towards the {+} side and normalized according to
\begin{equation}
			  n\: .\: n\: =\: \epsilon \hspace{1cm},
\end{equation}
where $\,\epsilon\,$ is constant over $\,\Sigma\,$ and 
takes respectively a positive, negative
or null value whenever $\,\Sigma\,$ is timelike, spacelike or lightlike. There
exists a non-vanishing smooth function $\, \chi \,$ on $\,\Sigma \,$
such that we have the relation
\begin{equation}
		n\,=\,{\chi}^{-1}\, \nabla \Phi \hspace{1cm}.
\end{equation}

	In order to deal with any type of hypersurface (timelike, spacelike
or lightlike) we introduce a vector field $\: N\:$, transversal to $\: \Sigma \: $
and satisfying
\begin{equation}
		    N\: .\: n\: =\: {\eta}^{-1}\hspace{1cm},
\end{equation}
where $\: {\eta}\:$ is a given non-vanishing smooth function on $\Sigma$.
 As $\,n\,$is normal to $\Sigma$, the vector
$N$ is defined up to a tangential displacement which
has to be the same on each side in order to make sure that the
same transversal vector is considered. For a timelike or spacelike hypersurface,
$\,N\,$ can be chosen identical to the normal $\,n\,$ 
and in that case one has $\,\epsilon\eta\,=\,1$ and $\eta$ is constant.
 
	For any functions $\,F^{\pm}\,$ defined in each domain $\,{\cal M}_{\pm}\,$,
we introduce the hybrid quantity
\begin{equation}
		\widetilde{F}\,=\,F^{+}\,\Theta (\Phi)+F^{-}\,\Theta(-\Phi)\hspace{1cm},
\end{equation}
where $\,\Theta \,$ is the Heaviside step-function ( which takes values $1$, ${1\over 2}$ or $0$
when its argument is respectively positive, null or negative )
and we write distributionally its derivative as
\begin{equation}
	{\partial}_{\mu} F\,=\,{({\partial}_{\mu} F)}^{\sim}
		+\chi \, n_{\mu}\, \lbrack\,F\,\rbrack\,\delta(\Phi)\hspace{1cm}.
\end{equation}
The metric ${\:g\:}$ and the scalar-fields $\,{\varphi}^{i}\,$ as also 
their tangential derivatives are continuous across
$\,\Sigma\,$ but their transverse derivatives are not and their jumps are
 defined by
\begin{equation}
 \lbrack\, {\partial}_{\mu} g_{\alpha \beta}\,\rbrack\,=\,\eta\,n_{\mu}\,{\gamma}_{\alpha \beta}
 \hspace{0.5cm},\hspace{0.5cm}
	\lbrack{\partial}_{\mu}{\varphi}^{i} \rbrack\,=\,\eta \, n_{\mu} \, {\zeta}^{i}\hspace{0.5cm},
\end{equation}
or using (21) by
\begin{equation}
   {\gamma}_{\alpha \beta}\,=\,N^{\mu}\,\lbrack\,{\partial}_{\mu} g_{\alpha \beta}\,\rbrack
    \hspace{0.5cm},\hspace{0.5cm}
    {\zeta}^{i}\,=\,N^{\mu}\,\lbrack\,{\partial}_{\mu}{\varphi}^{i}\,\rbrack 
	\hspace{1cm}.
\end{equation}

	It can be checked \cite{BI} that both the
$\,{\gamma}_{\alpha \beta}$'s and the
$\,{\zeta}^{i}$'s are independent on the choice of the transversal vector $N$.
Furthermore as only the projection of $\,{\gamma}_{\alpha \beta} \,$ 
onto $\, \Sigma \,$ has an intrinsic
meaning, they are not uniquely defined by the above equations and
one may perform the gauge transformation
\begin{equation}
   {\gamma}_{\alpha \beta}\rightarrow {\gamma}_{\alpha \beta}
		+ 2\,{\lambda}_{(\alpha} n_{\beta)} \hspace{1cm},
\end{equation}
where $\,{\lambda}_{\alpha}\,$ are the components of an arbitrary
vector-field over $\,\Sigma\,$.
Using the hybrid notation (22) for the metric $\,g\,$ and the scalar fields
$\,{\varphi}^{i}\,$, and the relations (24) for their derivatives, it can be shown that
the Einstein tensor and the d'Alembertian of the scalar fields are equal to
\begin{equation}
  G_{\mu \nu}\,=\,{G_{\mu \nu}}^{\hspace{-0.3cm}\sim}+
	\eta \, \chi \, \delta (\Phi)\,\left[\,{\gamma}^{(\,\mu}\, n^{\nu \, )}
	-\frac{1}{2}\,\left( \, \gamma n^{\mu} n^{\nu}+{\gamma}^{\dagger}\, g^{\mu \nu}
	+\epsilon \,\lbrack \, {\gamma}^{\mu \nu}
				-\gamma \, g^{\mu \nu} \, \rbrack \, \right)\,\right]
\end{equation}
\begin{equation}
	\Box{\varphi}^{i}\,=\,\left({\Box{\varphi}^{i}}\right)^{\sim}+
				\epsilon\,\eta \, \chi \, \delta(\Phi)\, {\zeta}^{i}\hspace{1cm}, 	
\end{equation}
where we have defined
\begin{equation}
{\gamma}_{\mu}\,=\,{\gamma}_{\mu \nu}\,n^{\nu} \hspace{0.5cm},
{\gamma}^{\dagger}\,=\,{\gamma}_{\mu}\,n^{\nu} \hspace{0.5cm},
\gamma \,=\,{\gamma}_{\mu \nu}\,g^{\mu \nu}\hspace{0.5cm}.
\end{equation}
We have used the fact that the product  
$\, \Theta (\Phi)\,\Theta(-\Phi)\,$ vanishes distributionally.

	Recalling that as a $\,\delta\,$-term has to appear in the
stress-energy tensor of the matter fields, one may write the 
total stress-energy tensor as
\begin{equation}
{\cal T}_{\mu \nu}\,=\,{\cal T}^{\sim}_{\mu \nu}+S_{\mu \nu}\,\chi\,\delta(\Phi)\hspace{1cm},
\end{equation}
where $\,S_{\mu \nu}\,$ represents the surface stress-energy tensor of the shell
and ${\cal T}^{\sim}_{\mu \nu}$ is of the form (22).
Then introducing
(27), (28) and (30) into the field equations (3) and (4),
and extracting their $\delta$-terms, one gets
\begin{equation}
	16\pi G\,{\eta}^{-1}\,S_{\mu \nu}\,=\,2{\gamma}^{(\,\mu}\, n^{\nu \, )}
	- \, \gamma \,n^{\mu} n^{\nu} - {\gamma}^{\dagger}\, g^{\mu \nu}
	-\epsilon \,\left( \, {\gamma}^{\mu \nu}
				-\gamma \, g^{\mu \nu} \, \right) \ ,
\end{equation}
\begin{equation}
	-4\pi G \,{\eta}^{-1}\,{\alpha}^{i}\,({\varphi})\,S\,=\,\epsilon\,{\zeta}^{i}\hspace{1cm},					
\end{equation}
where $\,S\,=\,S^{\mu \nu}\,g_{\mu \nu}\,$.

	The surface stress-energy tensor $\,S^{\mu \nu}\,$ keeps then the same
form as in general relativity, eq.(17) of \cite{BI}. 
In the Jordan-Fierz frame it would have
taken a different and more complicated form including the jumps
of the scalar fields. It can be checked from (31)
that $S^{\mu \nu}$ is a tangential quantity
\begin{equation}
			S^{\mu \nu}\,n_{\nu}\,=\,0 \hspace{1cm},
\end{equation}
and that it is invariant under the gauge transformation (26).
Eliminating the trace $\,S\,$ of $S_{\mu \nu}$ between 
(31) and (32), one obtains a relation
between the jumps $\,{\gamma}_{\mu \nu}\,$ and $\,{\zeta}^i\,$
\begin{equation}
{\alpha}^{i}\,({\varphi})\,({\gamma}^{\dagger}-\epsilon\gamma)\,=\,2\epsilon\,{\zeta}^{i}\hspace{1cm}.
\end{equation}
As $\:{\gamma}^{\dagger}-\epsilon\gamma\:$ is invariant under the gauge
transformation (26), this fulfils the intrinsic nature 
of the junction conditions in spite of the non unicity of the transversal 
$\:N\:$. Moreover, as one can see from (34), the jumps in
the first derivatives of the metric and the scalar fields 
cannot be choosen independently of each other, and this provides an extra
boundary condition for the resolution of the field equations.

Using again the hybrid notations (22-23), we obtain from the Bianchi identities 
the conservation relation of the generalized stress-energy tensor (30)
\begin{equation}
		{\nabla}_{\mu}\,{\cal T}^{\mu \nu}\,=\,0\hspace{1cm}.
\end{equation}
Introducing (20) and (30) in (35) one gets
\begin{equation}
	{\nabla}_{\mu}\,(\chi\,S^{\mu \nu})\,\delta(\Phi)\,
		=\,-{\nabla}_{\mu}\,{{\cal T}^{\mu \nu}}_{\hspace{-0.3cm}\sim}\hspace{1cm},
\end{equation}
and extracting the $\,\delta\,$-terms one obtains the equation of conservation  
for the surface stress-energy tensor of the shell
\begin{equation}
  {\nabla}_{\nu}(\,\chi\,S^{\mu \nu})\,=-\,\left[\,T^{\mu \nu}n_{\nu}\,\right]\,\chi\,
  	       	-\chi \, \lbrack\,T^{\mu \nu}({\varphi})\,\rbrack\,n_{\nu}\hspace{1cm},
\end{equation}
where $\,{\nabla}\,$ here stands for the covariant derivative operator associated
with the hybrid metric $\, g^{\sim}_{\mu \nu}\,$, 
i.e. such that $\,{\nabla}_{\mu}\,g^{\sim}_{\lambda \nu}=0\,$. The whole
set of equations (31), (32) an d (37) describe the dynamics of the
shell, however as they are not all independent only part of them
need to be used.
\vspace{0.5cm}

{\centerline{{\it Timelike and spacelike shells ($\,\epsilon\,\not= \,0\,$)} :}} 	
	For these hypersurfaces, the gauge freedom (26) enables us to choose
 $\,{\gamma}_{\mu \nu}\,$ such that its contraction with the normal vector
vanishes, hence $\,{\gamma}^{\mu}\,=\,0\,$ and $\,{\gamma}^{\dagger}\,=\,0\,$.		
With that choice the relations (31-32) reduce to
\begin{equation}
	-16 \pi G {\eta}^{-1}\,S_{\mu \nu}\,=\,\epsilon\,{\gamma}_{\mu \nu}
					-\gamma \, (\epsilon\,g_{\mu \nu}-n_{\mu} n_{\nu})
\end{equation}
\begin{equation}
	-4\pi G \, {\eta}^{-1}\,{\alpha}^{i}\,({\varphi})\,S\,=\,\epsilon\,{\zeta}^{i}
\end{equation}
where the last equation can also be replaced by (34), i.e. by
\begin{equation}
        {\gamma} {\alpha}^{i}({\varphi}) = -2 {\zeta}^i\hspace{1cm}.
\end{equation}
The equation (38) is the same as in general relativity but we have here the
additional constraints (39) or (40) coming from the presence of the
scalar fields $\: {\varphi}^{i} \:$.
\vspace{0.5cm}

{\centerline{{\it Lightlike shell ($\,\epsilon\,=\,0\,$) :}}} 
	In that case, it follows from (31) that the stress-energy tensor is equal to		
\begin{equation}
	-16 \pi G {\eta}^{-1}\,S^{\mu \nu}\,=\, \gamma \,n^{\mu} n^{\nu}
	-(\,{\gamma}^{\mu}\, n^{\nu}+{\gamma}^{\nu}\, n^{\mu}\,) \hspace{1cm},
\end{equation}
and from (32) that it is tracefree $\,S\,=\,0\,$, or equivalently from (34)
, $\,{\gamma}^{\dagger}\,=0\,$. This last property can be
interpreted as a condition for the shell to be presureless as one can show. 
At first the normal vector
$\,n\,$ is tangent to the null generators of the hypersurface $\,\Sigma\,$ and
satisfies the geodesic equation
\begin{equation}
		     {\nabla}_{n}\, n=\kappa \, n \hspace{1cm},
\end{equation}
where $\,\kappa\,$ vanishes whenever $\,n\,$ is associated to an affine
parametrization along the null generators. It follows from the definition of
$\,{\gamma}_{\mu \nu}\,$ that
$\,{\gamma}_{\mu \nu}\,=\,\left[\,{\cal L}_{N}\,g_{\mu \nu}\,\right]\,$,
then using $\,n.n\,=\,0\,$ and the properties of the Lie derivative
${\cal L}_{N}$, one successively gets
$$\,{\gamma}^{\dagger}\,=\,n^{\mu}\,n^{\nu}\,{\gamma}_{\mu \nu}\,=\,
	-2\,\left[\,n.{\cal L}_{N}\,n\,\right]\,=\,2\,\left[\,n.{\nabla}_{n}N\right]\,=\,
	-2\,{\eta}^{-1}\,\left[\,\kappa\,\right] \hspace{1cm} ,$$
where we have also used $\,\left[\,{\nabla}_{n}\,{\eta}\,\right]\,=\,0\,$.
On the other hand, taking the jump accross $\,\Sigma\,$ of the Raychauduri's
formula the null generators of $\Sigma$, one obtains
\begin{equation}
\lbrack{\kappa} \rbrack \, \Theta \,=
\, 8\pi G\,\lbrack{\,{\cal T}_{\mu \nu}\,n^{\mu}n^{\nu}}\,\rbrack \hspace{1cm},
\end{equation}
where $\,\Theta\,$ is the dilation rate of the null generators.
As the normal derivative of the scalar fields
$\,n^{\mu}\,{\nabla}_{\mu}{\varphi}^{i}\,$  is continuous on
$\,\Sigma\,$ -recall that the normal is tangent to a null hypersurface-
the scalar field part (5) of the total stress energy tensor vanishes in (43)  
and ${\cal T}_{\mu \nu}$ can be replaced by
the matter stress energy tensor $T_{\mu \nu}$ in this equation.
Finally combining these results one gets
\begin{equation}
{\gamma}^{\dagger}\,\Theta\,=\,-16\pi \, \eta\,\left[\,T_{\mu \nu}\,n^{\mu}n^{\nu}\,\right]\hspace{1cm},
\end{equation}
which shows that $\, {\gamma}^{\dagger} / {16\pi \eta}\,$ represents
the isotropic surface pressure of the null shell.
Hence, as $\: {{\gamma}^{\dagger}} = 0$, a null shell
cannot possess any isotropic pressure but only energy density 
and possibly shears - this restriction does not exist in general relativity
where a null shell can have a surface pressure, see examples in
\cite{BI}, \cite{BBH}.
Another consequence is that the vanishing of the right-hand side (44) 
implies that no energy can be transferred to the shell 
from the surrounding matter fields.

   It has been shown in general relativity \cite{BI} that in the lightlike case
a shell is generally accompanied by an impulsive wave. This property remains
unchanged in scalar-tensor theories and we briefly recall here why such
a decomposition into a shell and wave occurs. The Weyl tensor of the space time
${\cal M}_{+} \bigcup {\cal M}_{-}$ contains a $\delta$-term which using (24)
 and the notations (22-23) can be shown to be equal to -see the eq.(41) of \cite{BI}
\begin{equation}
C^{\alpha \beta}_{\ \ \mu \nu} = \lbrace \,2 \eta\,
n^{\lbrack\alpha} {\gamma}^{\beta\rbrack}_{\lbrack\mu}
n_{\nu\rbrack}
- 16 \pi \, {\delta}^{\lbrack\alpha}_{\lbrack\mu}
S^{\beta\rbrack}_{\nu\rbrack} 
 + \frac{8}{3} \pi \,S^{\lambda}_{\lambda}
{\delta}^{\alpha \beta}_{\mu \nu}\,\rbrace 
\,\chi\, \delta(\Phi) \hspace{1cm}.
\end{equation}
	It can be seen from (41) and (29) that only
$\gamma_{\mu \nu}n^{\nu}$and $\gamma$
enters for the expression (41) for $S^{\mu \nu}$, 
with the additional property $\gamma^{\dagger}=0$ in scalar-tensor theories.
Therefore there remains a part of $\gamma^{\mu \nu}$ which only contributes
to the first term of (45) and can be interpreted as being due to the
presence of an impulsive gravitational wave. 
As a shell and a wave generally co-exist the null
hypersurface $\Sigma$ is at the same time the history of a shell
and of a wave-front. More on this subject will be said in the next section,
and an example of this situation will be described
\addtocounter{Partie}{3}
in sect.{\Roman{Partie}}.
\\
\addtocounter{Partie}{-2}
\nssection{The intrinsic description}

	We denote by $\: {\xi}^{a}\:$ ($a\: =\: 1,2,3\: $) a set of 
intrinsic parameters for the hypersurface $\: \Sigma \: $, and
$\: e_{(a)}\: =\: {\partial}/{\partial {\xi}^{a}}\: $ the corresponding
tangent basis vectors. The induced metric $\,g_{ab}\,$ on the shell is then given by
\begin{equation}
		g_{ab}\,=\, e_{(a)} . e_{(b)}\hspace{1cm}.
\end{equation}
In the case of a timelike or spacelike shell, one needs to introduce
the extrinsic curvature $\,K_{ab}\: \equiv\: -n\: .\: {\nabla}_{e_{(b)}}e_{(a)}\,$
where $n$ is the unit normal, $\: {\epsilon} = \pm 1 $ and the shell is
characterized by the jump of $K_{ab}$ accross $\Sigma$.
For lightlike shells, this quantity does not carry any
extrinsic information because of the tangent nature of the normal vector $n$,
and we introduce the transverse extrinsic curvature
\begin{equation}
     {\cal K}_{ab}\: \equiv\: -N\: .\: {\nabla}_{e_{(b)}}e_{(a)}\:
		     =\: -N\: .\: \frac{\delta e_{(a)}}{\delta {\xi}^{b}}\hspace{1cm},
\end{equation}
where $N$ is the transversal vector already introduced in 
the previous section (21).
Althougth $\: {\cal K}_{ab}\:$ depends on the choice of the vector $\:N \:$
it has been shown in \cite{BI} that its jump across $\: \Sigma \: $ 
\begin{equation}
      {\gamma}_{ab}\: \equiv\: 2\: \left\lbrack{\cal K}_{ab}\right\rbrack\ \  ,
\end{equation}
is a well defined quantity which is free of the arbitrariness in the
transversal $N$. It can be shown \cite{BI} that $ {\gamma}_{ab} $ is the projection
onto $ \Sigma $ of the $ {\gamma}_{\mu \nu} $'s introduced in the eq.(24), i.e.
$ {\gamma}_{ab} = {\gamma}_{\mu \nu} e^{\mu}_{(a)} e^{\nu}_{(b)} $.

	The four vectors $\,(\,N\,,\, e_{(a)}\,)\,$ form an oblique
 basis whith respect to which
the normal vector n can be decomposed as
\begin{equation}
	       n\: =\: \epsilon \: \eta \: N + l^{a}\: e_{(a)} \hspace{1cm},
\end{equation}
where $\: l^{a}\:$ are smooth functions. It follows from this decomposition that
\begin{equation}
			g_{ab}\,l^{b}\,=\,-\epsilon\,\eta\,N_{a}\hspace{1cm},
\end{equation}
where $\,N_{a}\,\equiv\,N. e_{(a)}\,$.

	The induced metric is degenerate whenever the shell
is lightlike and in that case its inverse cannot be defined.
In order to generalize the notion of an inverse metric as a
raising indices operator valid in any case, we introduce as in \cite{BI} the symmetric
matrix $\,g_{*}^{ab}\,$ such that
\begin{equation}
	g_{*}^{ac}\,g_{cb}\,=\,{\delta}^{a}_{b}-\eta\,l^{a}\,N_{b}\hspace{1cm}.
\end{equation}
$\,g_{*}^{ab}\,$ is not uniquely defined by the above relation because
one may perform the following transformation
$\,g_{*}^{ab}\rightarrow g_{*}^{ab}+2\,\lambda\,l^a\,l^b\,$, where $\,\lambda\,$ 
is an arbitrary function, without changing (51). In the
non lightlike case ($\epsilon \not= 0$), 
a convenient choice is $\,N=n\,$, hence $\,N_a\,=\,0\,$,
and $\,g_{*}^{ab}\,$ is the usual inverse metric $\,g^{ab}\,$. In the lightlike case
($\,\epsilon\,=\,0\,$) , we have $\,l^a\,\not=\,0\,$ and $\,N_a\,\not=0$,
and it can be checked that $\,g^{ab}_{*}\,g_{ab}\,=\,2\,$.
The completeness relation which in the normal basis $(n,e_{(a)})$ is
\begin{equation}
g^{\mu \nu}\,=\,g_{*}^{ab}\, e_{(a)}^{\mu} \,e_{(b)}^{\nu}+{\epsilon}^{-1}\,n^{\mu}n^{\nu}\hspace{1cm},
\end{equation}
becomes in the oblique basis $\,(N,e_{(a)})\,$
\begin{equation}
g^{\mu \nu}\,=\,g_{*}^{ab}\, e_{(a)}^{\mu} \,e_{(b)}^{\nu}
		+2\eta l^{a} e_{(a)}^{(\mu}N^{\nu)}
		+{\eta}^{2}\epsilon\,N^{\mu}N^{\nu}\hspace{0.5cm}.
\end{equation}
Because of the tangential nature of the surface stress-energy 
of the shell $\,S_{\mu \nu}\,$, see (33), 
we can write 
\begin{equation}
	S^{\mu \nu}\,=\,S^{ab}\,e_{(a)}^{\mu} \,e_{(b)}^{\nu}\hspace{1cm},
\end{equation}
where $\,S^{ab}\,$ is now an intrinsic tensor of $\,\Sigma\,$.
As it is known from the distributional description that the surface-energy 
tensor keeps the same form as in general relativity, its intrinsic
form is still equal to (see the eq.(31) of ref.\cite{BI})
\begin{equation}
		16\pi G\, {\eta}^{-1}\,S^{ab}\,=\left(g_{*}^{ac}\,l^{b}l^{d}
		+l^{a}l^{c}\,g_{*}^{bd}-g_{*}^{ab}\,l^{c}l^{d}-l^{a}l^{b}\,g_{*}^{cd}
	   						\right){\gamma}_{cd}\,
		-\,\epsilon \,\left(\,g_{*}^{ac}g_{*}^{bd}-g_{*}^{ab}g_{*}^{cd}
							\, \right){\gamma}_{cd}	
\end{equation}
\vspace{0.5cm}
{\centerline{{\it Timelike and spacelike shells $\epsilon \not= 0$} :}
  Making the convenient choice such that $\,N=n\,$, one recovers 
from (55) the well-known relation
\begin{equation}
	16\pi\,G\,S^{ab}\,=\,-{\gamma}^{ab}+\gamma\,g^{ab}\hspace{1cm},
\end{equation}
where we have used $\,l^a\,=\,0\,$ , $\,\epsilon\,\eta\,=\,1\,$ ,
 $\,g_{*}^{ab}\,=\,g^{ab}\,$,and
$\,{\gamma}_{ab}\,=\,2\,\left[\,K_{ab}\,\right]\,$ .
The equation for the scalar field is again given by (39) or (40) where now
the trace $S$ is taken from (56).

The influence of the scalar fields $\,{\varphi}^{i}\,$ can also be seen in the
the hamiltonian and momentum constraints which yield here the two following equations
\begin{equation}
S_{a ; b}^{\; b}\, =\, -\,\left\lbrack T_{\mu \nu}\, e_{(a)}^{\mu}
					\, n^{\mu} \right\rbrack
		  -{{\gamma}_{ij}\over {4 \pi G}}\,{\zeta}^i\,{\nabla}_{a}{\varphi}^{j}		   
\end{equation}
\begin{equation}
S^{ab}\, {\tilde K}_{ab}\,=\, \left\lbrack T_{\mu \nu}\,n^{\mu}n^{\nu}
						\,\right\rbrack
		 +{{\gamma}_{ij}\over {8\pi G}}\left\lbrack
                                       {\nabla}_{n}{\varphi}^{i}
		                       {\nabla}_{n}{\varphi}^{j}
                                                \right\rbrack \hspace{1cm},
\end{equation}
where ; is the covariant differentiation with respect to $\,g_{ab}\,$,
${\nabla}_{n} = n^{\mu}{\nabla}_{\mu}$ is the normal derivative, and the
tilde is the average $\,{\tilde F}\,=\,(F^++F^-)/2\,$.
\vspace{0.5cm}

{\centerline{{\it lightlike shells $\epsilon = 0$} :}
	In that case, we have $\,g_{ab}\,g^{ab}_{*}\,=\,2\,$ and from (50) 
$\,g_{ab}\,l^b\,=\,0$. 
The trace-free (or pressureless) property which was obtained in the distributional
description corresponds here to
\begin{equation}
	{\gamma}^{\dagger}\,=\,{\gamma}_{cd}\,l^c\,l^d\,=\,0 \hspace{1cm},
\end{equation}
thus giving for the surface stress-energy tensor (55)
\begin{equation}
	16\pi G\, {\eta}^{-1}\,S^{ab}\,=\left(g_{*}^{ac}\,l^{b}l^{d}
	+l^{a}l^{c}\,g_{*}^{bd}-l^{a}l^{b}\,g_{*}^{cd}
	 				\right){\gamma}_{cd}\hspace{1cm}.
\end{equation}
As already mentioned in the distributional description a shell and a wave
generally co-exist in the lightlike case. The part of $\gamma_{ab}$ which
only enters the expression (60) of $S^{ab}$ is $\gamma_{ab} l^{b}$ and
$\gamma = g^{cd}_{*} \gamma_{cd}$. This leaves two independent components,
denoted by $\hat{\gamma}_{ab}$, corresponding to an impulsive gravitational
wave and related to the two degrees of freedom of polarization of the wave.
The expression of $\hat{\gamma}_{ab}$ is \cite{BBH}
\begin{equation}
 \hat{\gamma}_{ab} = \gamma_{ab} - \frac{\gamma}{2} g_{ab} + \eta ( N_{a} \gamma_{bc} + N_{b} \gamma_{ac} ) l^c\hspace{1cm},
\end{equation}
where we have used (59).
 
\addtocounter{Partie}{1}
\nssection{Existence of discontinuous gauge fields}

	When gauge fields are present, the stress-energy tensor
$T_{\mu \nu}$ which appears in the total stress-energy tensor 
${\cal T}_{\mu \nu}$, see the eq.(17), is the sum of a pure matter part 
$T^{\mu \nu}_{m}$ and a gauge field part $T^{\mu \nu}_{F}$ i.e.
$T^{\mu \nu}=T^{\mu \nu}_{m}+T^{\mu \nu}_{F}$. Therefore, if there exists a 
thin shell it will in general carry surface charges and currents
acting as sources for the gauge fields and producing discontinuities
in these fields accross the shell. The surface charges and currents will then enter 
the 4-vector current in the form of a Dirac $\delta$-term which is related
to the discontinuities of the gauge fields in the same way as the surface 
stress-energy tensor $S^{\mu \nu}$ is related to the discontinuity 
of the first derivatives of the metric tensor.

	Let us consider the general case of a non abelian gauge field 
of the Yang-Mills type 
\begin{equation}
        {F}^{a}_{\mu \nu}={\nabla}_{\mu} A^{a}_{\nu}-
                                {\nabla}_{\nu} A^{a}_{\mu}
                                -e(A_{\mu} \wedge A_{\nu})^a \hspace{1cm}.
\end{equation}
In this section only, the latin indices refer to the gauge group and
cannot be confused with the parameters of the hypersurface introduced earlier. 
We still use the distributional notation of 
sect. {\Roman {Partie}}.1 . In order to produce a surface current, the potential
vector $A^{a}_{\mu}$ must be only $C^0$ on the hypersurface $\Sigma$, i.e.  
$[A^{a}_{\mu}]=0$ and one can write the jump of its first derivatives 
accross $\Sigma$ as
\begin{equation}
	[{\partial}_{\mu} A^{a}_{\nu}]=\eta n_{\mu} {\lambda}^{a}_{\nu}\hspace{1cm},
\end{equation}
where ${\lambda}^{a}_{\nu}$ is a vector field defined on $\Sigma$ only 
-see below for more on $\lambda^{a}_{\nu}$.
The corresponding gauge field is dicontinuous accross $\Sigma$ 
and using (63) one gets
\begin{equation}
	[{F}^{a}_{\mu \nu}]=\eta ( n_{\mu} {\lambda}^{a}_{\nu}-
				 n_{\nu} {\lambda}^{a}_{\mu})
				\hspace{1cm}.
\end{equation}
Using this result and the hybrid notation of sect. {\Roman{Partie}}.1,
one gets for the Yang-Mills field equations
\begin{equation}
{\nabla}_{\nu} {\tilde F}^{\mu \nu}_a = 4\pi [{\tilde J}^{\mu}_a+
				j^{\mu}_a \chi \delta(\Phi)]\hspace{1cm},
\end{equation}	
where ${\tilde J}^{\mu}_a=J^{\mu +}_a \Theta(\Phi)+
					J^{\mu -}_a \Theta(-\Phi)$
represents the 4-current in the domains ${\cal M}_{\pm}$ and
$j^{\mu}_a$ is the surface current. Identifying the $\delta$-terms of
each side of (65) one gets
\begin{equation}
	4\pi j^{\mu}_{a} = \eta ({\lambda}_{a} . n) n^{\mu}
				-\epsilon \eta {\lambda}^{\mu}_{a} \hspace{1cm}. 
\end{equation}
It can be checked from this expression that $j^{\mu}$ is a tangential
quantity, $j.n=0$.

	The vector field ${\lambda}^{a}_{\mu}$ which has been introduced in 
(63) is not uniquely determined by this equation. As it must have a unique 
projection onto the
hypersurface $\Sigma$, it is in fact only defined up to the transformation
\begin{equation}
	{\lambda}^{a}_{\mu} \rightarrow  {\lambda}^{a}_{\mu}+C^a n_{\mu}
	\hspace{1cm},
\end{equation}
where $C^a$ is an arbitrary spacetime scalar. In the case of 
a timelike or spacelike shell ($\epsilon \not= 0$) one can use (67) to 
choose $C^a$ in order that ${\lambda}^{a}_{\mu}$ is purely tangent
, i.e. $\lambda_{a} .n=0$, and the surface current (66) reduces to
\begin{equation}
		4\pi j^{\mu}_{a}\ =\ -\epsilon \eta {\lambda}^{\mu}_{a}\hspace{1cm}.
\end{equation}
For a lightlike shell ($\epsilon = 0$) such a choice cannot be
done and one has
\begin{equation} 
	4\pi j^{\mu}_{a} \ =\ \eta (\lambda_{a} .n) n^{\mu} \hspace{1cm}.
\end{equation}
Hence only the part $\lambda_{a}.n$ of the vector $\lambda_a$ contributes
to the surface current. The remaining part $\hat{\lambda_a}$ can be written as
\begin{equation}
     \hat{\lambda^{\mu}_{a}} \,=\, \lambda^{\mu}_{a}-(\lambda_{a}.n)N^{\mu}\hspace{1cm},
\end{equation}
and characterizes the shock wave which is associated with the discontinuity
of the gauge field.
	Finally, let us consider the behaviour of the stress-energy
tensor $T_{\mu \nu}$. Its gauge field part $T^{\mu \nu}_{F}$ has the general form
\begin{equation}
	 T^{\mu \nu}_{F} =  F^{\mu \lambda}_{a} F^{a \nu}_{\lambda} 
	 - \frac{1}{4}F^2 g^{\mu \nu} \hspace{1cm}.
\end{equation}
According to (64) it is discontinuous across $\Sigma$, 
i.e. $[T^{\mu \nu}_{F}]\not= 0$, and therefore
the $\delta$-term which is necessary for the existence of
a shell can only come from the pure matter part $T^{\mu \nu}_{m}$.\\

\addtocounter{Partie}{1}
\addtocounter{sPartie}{1}

\nsection{SPHERICALLY SYMMETRIC SHELLS}

	The first example of a thin shell in scalar-tensor theories that we consider
is a spherical bubble separating two domains $\,{\cal M}_{\pm}$ 
where the metrics $\,g^{\pm}_{\mu \nu}\,$
and the scalar fields $\,{\varphi}^{i}_{\pm}\,$ are 
spherically symmetric. Because of the
presence of the scalar fields, the Birkhoff's theorem no longer apply and the exterior
vacuum solution, i.e. $\,T_{\mu \nu}\,=\,0\,$ in (3),  
is not necessarly static. For simplicity, we shall
here restrict ourselves to static solutions and  
use the following form of the metrics in the two domains
\begin{equation}
	ds^{2}_{\pm}\,=\,-f_{\pm}(R_{\pm})\,e^{2{\psi}_{\pm}(R_{\pm})}\,dt^{2}
	+f^{-1}_{\pm}(R_{\pm})\,d{R_{\pm}}^{2}+r^{2}_{\pm}(R_{\pm})\,d{\Omega}^{2}\hspace{0.5cm},
\end{equation}
where $\,d\Omega ^2 \,=\,d\theta ^2+{\sin}^2 \theta \, d \phi ^2 \,$ 
is the spherical line element.
The scalar fields $\,{\varphi}^{i}_{\pm}$ and the three 
functions $\,f\,$, $\,\psi\,$, and $r$ only depend on the coordinate $\,R\,$.
Some particular solutions to the field equations in static spherically
symmetric spacetimes are presented in appendix A.

	As the induced metric and the scalar fields are continuous 
on the hypersurface  ${\Sigma}$  corresponding to the shell we must have
on $\Sigma$
\begin{equation}
	\left\lbrace
		\begin{array}{lll}
			r_{+}(R_+) & = & r_{-}(R_-)\\
			{\varphi}^{i}_{+}(R_+) & = & {\varphi}^{i}_{-}(R_-)\hspace{1cm}.	
		\end{array}
	\right.				
\end{equation}
These matching conditions limit the evolution of the shell and the nature of the
junction. Unless they are trivially satisfied, as it may happen if the
metrics $\,g^{\pm}_{\mu \nu}\,$ and the scalar fields $\,{\varphi}^{i}_{\pm}\,$
are identical, the above equations imply that $\,R_{\pm}\,$ take
constant values and the shell has to be stationary (examples of
these two situations will be given later on). In
general relativity only the first equation (73) is present and the shell is 
not necessarily static but can have a radial motion. 
It also follows from (73) that in the case of a lightlike shell it 
can only be located on a common horizon of the two spacetimes
$\,{\cal M}_{\pm}\,$ as one must have a null and stationnary hypersurface. 
However as it is known that no black hole solution with regular
event horizons exist in scalar-tensor theories, no lightlike shell
can be introduced - the situation is
different with dilaton theories where black hole solutions 
exist (see for instance \cite{Hor}).
The fact that the junction can only be made on a
stationary hypersurface is actually a consequence of 
our assumption of staticity. Had we
removed this condition and considered non static spherically symmetric spacetimes
(for instance anologous to the Vaidya solution), 
radially expanding or contracting shells
might have been introduced.

	For simplicity we shall henceforth assume that only one scalar field is
present. Let us consider a timelike shell and let it  for the moment
have an arbitrary radial motion. The induced metric 
on the timelike surface $\,\Sigma\,$ takes the form
\begin{equation}
		ds^2\,=\,-d\tau ^2+r^2 (\tau)\,d\Omega ^2\hspace{1cm},
\end{equation}
where $\,\tau\,$ is the proper time.
The normalized velocity $\,u\,=\,d/d{\tau}\,$ ($\,u.u\,=\,-1\,$)
and the unit normal $\,n\,$ ($\,n.n\,=\,1\,$, $\,u.n\,=\,0\,$) 
have components given by (omitting the $\pm$ indices)
\begin{equation}
	u^{\alpha}\,=\,\left(\frac{{\epsilon}_1 \sqrt{f+\dot{R}^{\, 2}}}{f\,e^{\psi}}\, ,
			\,\dot{R}\, ,\,0\, , \,0\, \right)
			\hspace{0.5cm},\hspace{0.5cm}
	n^{\alpha}\,=\,\left(\frac{{\epsilon}_2 \,\dot{R}}{f\,e^{\psi}}\, ,
		\,{\epsilon}_1{\epsilon}_2\sqrt{f+\dot{R}^{\, 2}}\, ,\,0\, , \,0\, \right)
\end{equation}
where  $ \dot{ } = d/d{\tau}$ , $\,{\epsilon}_1\, , \,
{\epsilon}_2\,=\,\pm 1\,$; note that
$\,{\epsilon}_1 \, {\epsilon}_2\,=\,$sign$\,(n^{\alpha}\,{\partial}_{\alpha}r)\,$.

	Because of the spherical symmetry, the surface stress-energy tensor
has the perfect fluid form
\begin{equation}
		S_{ab}\,=\,(\sigma+p)\,u_a u_b + p\,g_{ab}\hspace{1cm},
\end{equation}
$\,\sigma\,$ being the surface energy density and $\,p\,$ the surface pressure.\\
Using the results of section \addtocounter{Partie}{-2}{{\Roman {Partie}}.1,2 , 
one obtains, as $N = n$ and $\epsilon = \eta = 1$
\begin{equation}
       -4\pi G\,\sigma \,=\,\left[\,K^{\theta}_{\hspace{0.2cm}\theta}\,\right]
\end{equation}
\begin{equation}
	8\pi G \, p \,=\,\left[\,K^{\tau}_{\hspace{0.2cm}\tau}\,\right]
		+\left[\,K^{\theta}_{\hspace{0.2cm}\theta}\,\right]
\end{equation}
\begin{equation}
	4\pi G \,(\sigma - 2p)\,{\alpha}^i \,= \,{\zeta}^i\hspace{1cm}.
\end{equation}
The non-zero components of the extrinsic curvature $ K_{ab} $ are equal to
\begin{equation}
   K^{\theta}_{\theta} = K^{\varphi}_{\varphi} = {\epsilon}_{1}{\epsilon}_{2} \frac{r'}{r} \sqrt{f+\dot{R}^{2}}
\end{equation}
\begin{equation}
   K^{\tau}_{\tau} = \frac{{\epsilon}_{1}{\epsilon}_{2}}{\sqrt{f+\dot{R}^{2}}} [\ddot{R} + \frac{f'}{2} + \frac{{\psi}'(f+\dot{R}^{2})}{2 e^{\psi}}]\hspace{1cm},
\end{equation}
with  $ \, ' = d/dR $.
Furthermore the $\tau$-component of the conservation equation (57) gives
\begin{equation}
\frac{dM}{d\tau}\,=\,-p\,\frac{dA}{d\tau}+A\left[\,T_{\mu \nu}\,u^{\mu}\,n^{\nu}\,\right]
			+ \frac{A\,{\zeta}^i}{4\pi G}\,{\dot \varphi}^j \, {\gamma}_{ij}\hspace{1cm},
\end{equation}
with $\,A\,=\,4\pi\,r^2 \,$ and $\,M\,=\,\sigma \,A\,$ being resp. the proper area 
and the inertial mass of the shell.
In the particular case of a domain wall, 
the equation of state is $\,\sigma +p\,=\,0\,$ and (82) becomes
\begin{equation}
	{\dot \sigma}\,=\,\left[\,T_{\mu \nu}\,u^{\mu}\,n^{\nu}\,\right]
			+3\,\sigma\,{\alpha}_i \, {\dot \varphi}^i\hspace{1cm}.
\end{equation}

The first example that we consider is a static spherical shell carrying an electric
charge $Q$. The interior spacetime $\,{\cal M}_-\,$ is flat with a metric
given by
\begin{equation}
     ds^{2}_{-}\,=\,-dt^{2}_{-} + dR^{2}_{-} + R^{2}_{-}d{\Omega}^2\hspace{1cm},
\end{equation}
and a constant scalar field, ${\varphi}_- = {\varphi}_0$. The exterior
spacetime corresponds to the Reissner-Nordstr\"{o}m-Brans-Dicke solution
described in appendix A. The shell is static and has
a constant radius $r_0$ which, according to the matching relations (73), 
satisfies $r_0 = R_{-0} = r_+(R_{+0})$.
Using these relations and the solution of appendix A one gets 
\begin{equation}
 r^{2}_{0} = e^{2(a-b){\varphi}_0/d} {[ch^2{\lambda} - e^{2b{\varphi}_0/d}sh^2{\lambda}]\over[1 - e^{2(a-b){\varphi}_0/d}]^2}\hspace{1cm}.
\end{equation}
where $\lambda$ is the charge parameter.
This equation shows that the constant radius of the shell depends
on the parameters of the spacetimes bordering the shell. The case of an
uncharged shell is simply obtained by putting ${\lambda}$ equal to zero. Finally
the surface energy density and pressure are obtained from (77) and (78) with
$\dot{R} = \ddot{R} = 0$ and the jump for the scalar field is given by (79).

  Our next example in spherical symmetry is a spherical domain wall
separating two identical vacuum spacetimes.
The metric and the scalar fields are still of the form (72). 
As the spacetimes are identical the matching relations (73) are
trivially satisfied and the shell can radially expand or contract.
For a domain wall we have $ \sigma + p = 0 $, and if it is assumed to be embedded
in either true or false vacuum  we have $\, T^{\pm}_{\mu \nu} u^{\mu} n^{\nu} = 0 \,$.
It then follows from (83) that the surface energy density
varies according to
\begin{equation}
     {\sigma} = {\sigma}_{0} e^{3 \alpha \varphi}\hspace{1cm}.
\end{equation}
where $ {\sigma}_0 = const$. It should  be noticed that the physical surface 
energy density is not $\sigma$ but rather the quantity ${\tilde \sigma}$ 
expressed in the Jordan-Fierz frame. If one uses the 
conformal transformation (15) one obtains the constant value 
${\tilde \sigma} = {\sigma}_0$
as expected for a domain wall embedded in true or false vacuum.

	The equation of motion of the domain wall is deduced from (77) with
$K^{\theta}_{\theta}$  given by (80). As the two sides of the wall are
identical, one sees that the shell only exists if the two products 
$({\epsilon}_1 {\epsilon}_2)_{\pm}$ take opposite values,
which corresponds to the situation where the shell separates either
two interior or two exterior geometries. Squaring (77) one gets the
following equation of motion
\begin{equation}
   {\dot R}^2 + V_{eff}(R) = -1\hspace{1cm},
\end{equation}
where we have introduced the effective potential
\begin{equation}
 V_{eff}(R) = -1 + f(R) - 4{\pi}^2G^{2}{\sigma}^2(R)r^{3}(R){(dr/dR)}^{-2}\hspace{0.5cm}.
\end{equation}
This equation shows that the motion is only possible provided
that $ V_{eff}(R) < -1$. If the spacetimes correspond to the true vacuum 
with a vanishing mass parameter b -see the appendix A, it can be shown
that the domain wall can only undergo a bouncing
motion starting from infinity down to some minimal radius and 
back to infinity. If $b$ does not vanish all types of motion
are a priori possible (monotonic and bouncing) according to the
values taken by the spacetime parameters $a,b$ and $d$ and the Brans Dicke
parameter $\alpha$. The situation is similar if we consider
false-vacuum instead of true-vacuum.\\

\addtocounter{Partie}{2}

\nsection{PLANAR SHELLS AND IMPULSIVE WAVES}
	
	The gravitational properties of planar shells have been extensively
studied in general relativity with particular emphasis on domain walls and
application to cosmology. The vacuum reflection-symmetric solution of the
Einstein equations for an infinitely thin planar domain wall was obtained
by Vilenkin \cite{Vil} and later generalized to walls with a given equation of
state \cite{IpSiLe} and without reflection symmetry \cite{WaLe}. 
Some planar solutions for supersymmmetric walls including
dilaton were also obtained by Cvetic et al \cite{Cv} and by Schmidt and Wang \cite{SW}
in the Brans Dicke theory.

         In this section, we present two examples with planar symmetry 
in a scalar-tensor theory where for simplicity only one scalar 
field $\varphi$ is introduced. We first give the exact solution for a domain wall 
surrounded by vacuum which is the counterpart 
in the Brans Dicke theory of the solution given by Vilenkin in general relativity. 
Then we study, as an illustration
of our formalism for the null case, a plane lightlike shell 
accompanied by a plane impulsive wave.  

	It is known from Taub \cite{Taub} that any 
plane-symmetric metric can be written as
\begin{equation}
	ds^2 \, = \,e^{2 \,\nu (t, z )}\,\left( -dt^2+dz^2 \right)
				+e^{2\, \mu (t,z )}\, \left( dx^2+dy^2 \right)
	\hspace{1cm}.
\end{equation}
where the plane of symmetry is $z=0$. Any reflection symmetric solution 
with respect to the plane $z=0$, where the domain wall is located, 
must satisfy $\nu (t, z )\, =\, \nu (t,-z )$ 
and $\mu (t, z )\, =\, \mu (t,-z )$, and for the scalar field 
$\varphi (t, z )\, =\, \varphi (t,-z )$. 
We shall henceforth call ${\nu}_0 (t)$, 
${\mu}_0 (t)$ and ${\varphi}_0 (t)$ the values of $\nu$, $\mu$ and $\varphi$ at the 
hypersurface $\Sigma$ , $z=0$. In this example the distributional 
description of section \addtocounter{Partie}{-3}{\Roman{Partie}}.1 will be used. The induced metric on $\Sigma$ is
\begin{equation}
ds^2 |_{\Sigma}\,=\,-e^{2{\nu}_0 (t)}\,dt^2+e^{2{\mu}_0 (t)}\,(dx^2+dy^2)
\hspace{1cm}.
\end{equation}
The unit normal is $n^{\alpha}=(0,e^{-{\nu}_0 },0,0)$ which from (20) implies 
that $\chi = e^{{\nu}_0}$, and $ \epsilon = +1 $ 
as we are in the timelike case. Choosing
$N=n$ for the transversal we have $\eta = 1$ in (21). Then one finds for 
the jumps ${\gamma}_{\alpha \beta}$ and $\zeta$ of the first derivatives 
of the metric and the scalar field which were defined in (24-25) 
\begin{equation}
	\left\lbrace
		\begin{array}{llllll}
{\gamma}_{00} & = & -{\gamma}_{11} & = & - 2\,e^{ {\nu}_0} [\,{\nu}_{,z} \,]\\
{\gamma}_{22} & = &  {\gamma}_{33} & = & e^{2{\mu}_0-{\nu}_0}\,\left[\,{\mu}_{,z}\, \right]\hspace{1cm},
		\end{array}
	\right.				
\end{equation}
and 
\begin{equation}
	\zeta \,= \,e^{-{\nu}_0} \,\left[\,{\varphi}_{,z} \right]\hspace{1cm}.  
\end{equation}
As the equation of state of the domain wall is $\sigma + p = 0$ one derives
from the equations of junction (38)
\begin{equation}
  4\pi G\, \sigma  =  -e^{-{\nu}_0}\,\left[\,{\mu}_{,z}\, \right]
\end{equation}
\begin{equation}
        \left[\,{\nu}_{,z}\, \right] \ =\ \left[\,{\mu}_{,z}\, \right]\hspace{1cm},
\end{equation}
and from the scalar field junction condition (39) 
\begin{equation}
\left[\,{\varphi}_{,z} \right] \,=\,-3 \alpha \, \left[\,{\mu}_{,z}\, \right]\hspace{1cm}.
\end{equation}
The time component of the equation of conservation (37) gives
\begin{equation}
{\partial\sigma \over \partial t}\,=\,-e^{ {\nu}_0 } \left[T^{\mu \nu} u^{\mu}n^{\nu} \right]
				+3 \alpha \sigma {\partial\varphi \over \partial t}\hspace{1cm}.  
\end{equation}
and as the domain wall is surrounded by vacuum on each side 
($T^{\pm}_{\mu \nu}=0$), it immediately follows that
\begin{equation}
		\sigma= {\sigma}_0 \, e^{3\alpha \, {\varphi}_0 (t)}\hspace{1cm},
\end{equation}
where ${\sigma}_0$ is a constant. It can be checked that the physical surface
energy density $\tilde{\sigma} $, which is derived from $\sigma$ by
using the conformal transformation (15) is a constant,
$\tilde{\sigma} = {\sigma}_0$, as expected for a domain wall in vacuum.

	The unknown functions $\mu (t,z)$, $\nu (t,z)$ of the metric
and the scalar field $\varphi (t,z)$ are derived from the field equations 
and the boundary conditions. Gathering all these results it can 
be shown that the exact solution for the metric and the scalar field is 
\begin{equation}
	ds^2\,=\,e^{a\,(9\,{\alpha}^2\,t-z)}\,(-dt^2+dz^2)+e^{a\,(t-z)}\,(dx^2+dy^2)
\end{equation}
\begin{equation}
		\varphi (t,z)\,=\,-\frac{3\,\alpha \, a}{2} \,(t-\mid z \mid)\hspace{1cm},
\end{equation}
where $a$ is a constant of integration which using (90) and (94), is related 
to ${\sigma}_0$ according to, $4\pi G\, {\sigma}_0 \,=\,a$.
The Vilenkin solution is recovered by putting the Brans-Dicke parameter
$\alpha$ equal to zero. As in this solution
the space-time is still locally flat 
everywhere except at $z=0$ and the $(t,z)$ part of the metric 
can be written in a form which is conformally related to the Rindler metric.

    The following example corresponds to the lightlike case.
It will be worked out in the intrinsic formalism developped in section 3.2.
An appropriate form of the metric is the Szekeres one
\begin{equation}
 ds^2 = -2 e^{-M}dudv + e^{-U} (e^V dx^2 + e^{-V} dy^2)\hspace{1cm},
\end{equation}
where the functions $M,U,V$ and the scalar field $\varphi$ only
depend on the null coordinates $(u,v)$, and $(x,y)$ are coordinates in
the planes $z=const.$ - we use the ordering $u,v,x,y$ and
greek indices range from $0$ to $3$.

	The null hypersurfaces, $u=const.$, are generated by null
geodesics with tangent, $n = \partial \ / \partial v$ - note that
$v$ is not an affine parameter. These null generators
 have expansion $\rho$ and shear $\sigma$ equal to 
\begin{equation}
\rho = -\frac{U_v}{2},\quad     \sigma = \frac{V_v}{2}\hspace{1cm},
\end{equation}
where the subscript $v$ indicates partial differentiation with respect
to $v$ (similar results hold on the hypersurfaces $v=const.$).

	The spacetime is divided at $\Sigma \,(u=0)$ into two halves $\cal M_+$ and
$\cal M_-$ where $\cal M_+$ ($\cal M_-$) is to the future (past) of
$\Sigma$ and corresponds to $u >0$ ($u <0$). To save subscripts we shall
drop the minus subscripts on any quantity refering to $\cal M_-$.
We assume that $\cal M_-$ is flat with coordinates ($u,v,x,y$) and
line element of the form (100) with $M=U=V=0$, and the scalar field is
constant $\varphi \equiv {\varphi}_0$. In the second half $\cal M_+$ the
coordinates are ($u,v_+,x_+,y_+$), the line element $ds^2_+$ is of the form (100),
with $M^+,U^+,V^+$ and the scalar field $\varphi^+$ depending
on the null coordinates $(u,v_+)$. We have taken for simplicity $u_+ = u$.
The two spacetimes are glued along $\Sigma$ by making the identification
\begin{equation}
 (0,v_+,x_+,y_+) = (0,v - F(x,y),x,y)\hspace{1cm},
\end{equation}
where $F$ is an arbitrary smooth function of $x$ and $y$ alone and produces
a shift in the null coordinate tangent to the hypersurface.

  We take $\xi^a = (v,x,y)$ with $a = 1,2,3$, as intrinsic parameters on $\Sigma$,
and as the normal is $n = e_{(1)}$ we obtain from (49) $l^a = \delta^a_1$.
Continuity at $u = 0$ of the induced metric
and of the scalar field requires that
\begin{eqnarray}
& &  U^{+}(0,v_{+}) = V^{+}(0,v_{+}) = 0,\\    
& &\varphi^{+}(0,v_{+}) = \varphi_0\hspace{1cm}.
\end{eqnarray}
The induced metric reduces to $\,g_{ab} = diag(0,1,1)$ and one may
take for its 'inverse' (51), $\,g^{ab}_* = diag(0,1,1)$.
A convenient choice for the transversal $N$ corresponds to $N.n = -1$, 
$N.e_{(2)} = N.e_{(3)} = 0$, $N.N = 0$, thus leading to components equal to 
$N^\alpha = (1,0,0,0)$ in $\cal M_-$, and 
\begin{equation}
N^{\alpha}_+ = (-e^{M^{+}_{0}}, \frac{F^2_x + F^2_y}{2}, -F_x, -F_y )\hspace{0.5cm},
\end{equation}
in $\cal M_+$ where $M^{+}_{0} \equiv M^{+}(0,v_{+})$.
Introducing these results into (47) one obtains the values of   
the transverse extrinsic curvature ${\cal K}_{ab}$ on each side of the the shell. 
As it  vanishes in $\cal M_-$ one simply gets from (48) 
for the jumps, $\gamma_{ab} = 2{\cal K}^{+}_{ab}$. 
Then using the trace-free property (59) one shows that
$M^+_0 = const.$ which must be equal to zero as the other side is flat, 
and the non-zero components of $\gamma_{ab}$ are equal to
\begin{eqnarray}
\nonumber & & \gamma_{22} = -2F_{xx} -U^{+}_{u}(0,v_+) + V^{+}_{u}(0,v_+) \\
\nonumber & & \gamma_{33} = -2F_{yy} -U^{+}_{u}(0,v_+) - V^{+}_{u}(0,v_+) \\
\nonumber & & \gamma_{23} = \gamma_{32} = -2F_{xy}\hspace{1cm}.
\end{eqnarray}
Therefore the surface stress-energy tensor (60) is of
the form, $- S^{ab} = \sigma l^a l^b $, and has only one
non-vanishing component equal to
\begin{equation}
  16 \pi G S^{11} = \gamma_{22} + \gamma_{33}
\end{equation}
These results show that the shell has no shear and is only characterized by its surface 
energy density which has the following expression in terms of the
functions $F$ and $U^+$
\begin{equation}
     8 \pi G \sigma = \Delta F + U^{+}_{u}(0,v_{+})\hspace{1cm}.
\end{equation}

On the other hand the wave part (61) of $\gamma_{ab}$ has the only
non-vanishing components
\begin{eqnarray}
& & \hat\gamma_{22} = -\hat\gamma_{33} = -F_{xx} +F_{yy} +V^{+}_{u}(0,v_{+}) \\
& & \hat\gamma_{23} = -2F_{xy}\hspace{1cm}.
\end{eqnarray}
According to the form taken by the functions 
$F(x,y)$, $U^{+}(u,v_{+})$ and $V^{+}(u,v_{+})$
diferent types of situations can occur:
we can have only a lighlike shell, only an impulsive wave or both.
A more extended version describing the geometry of the spacetimes
$\cal M_{\pm}$ and the properties of the shell and the wave
will be presented in forthcoming paper \cite {Bres}.\\

\addtocounter{Partie}{3}

\nsection{CONCLUDING REMARKS}

    Besides the scalar-tensor theories of gravity which have here been
considered there exist, as mentionned in the introduction,
other alternative theories of gravity which
also introduce scalar fields such as the dilaton. 
Although they look quite similar to scalar-tensor theories when no matter
field is present, the dilatonic theories of gravity present an important
difference which is due to the way the dilaton couples to the other fields.
In this last section we would like to briefly discuss the smoothness
properties of the dilaton accross a singular hypersurface $\Sigma$
and point out how they differ from those obtained in the scalar-tensor theories.

  Let us use the following expression for the action in the presence of a 
dilaton in the Einstein metric \cite{Da}
\begin{equation}
S=\int d^4 x \frac{\sqrt{-g}}{4 q}\left[ R
        -2 \left(\nabla \varphi\right)^2\right]
        -\int d^4 x \frac{\sqrt{-g}}{4} k_F \, e^{-2 \kappa \varphi} F^2
       +\,S_{m}\left[ \Psi,\varphi,g\right]\, ,
\end{equation}
where $\varphi$ is the dilaton, $\Psi$ a matter field, $F$ a
Maxwell field with $F=dA$, $A$ being the potential.
Here, $q$ is the gravitational coupling constant ($q=4\pi G$),
$\kappa$ is the coupling constant to the dilaton, and $k_F$
is the coupling constant for the gauge field $F$. The action $S_m$
for the matter fields $\Psi$ can for instance be taken as
\begin{equation}
        S_{m}\left[\Psi,\varphi,g^{\alpha \beta}\right]=
        \int d^4 x \sqrt{-g} \left[ -\frac{1}{2}\left(D_{\alpha}\Psi\right)
        \left(D^{\alpha}\Psi\right)^{\star}-e^{2 \kappa \varphi}V(\Psi)\right]
        \hspace{1cm},
\end{equation}
where $D_{\alpha}={\partial}_{\alpha}+ieA_{\alpha}$ is the gauge-covariant derivative,
$e$ the associated charge, and $V$ a potential.

        Such an expression for the total action $S$ shows that 
the dilaton does not minimally (i.e. metrically) couple to the
different fields but that it induces spacetime dependent coupling factors.
The field equation for the dilaton which follows from this action is
\begin{equation}
        \Box \varphi = -\frac{q \kappa k_F}{2} e^{-2 \kappa \varphi} F^2
                                  +2 q \kappa e^{2 \kappa \varphi} V(\Psi)\hspace{1cm}.
\end{equation}

In comparison with the analog equation (4) in the scalar-tensor theories, one
immediately sees that only the potential $V(\Psi)$ and not the trace $T(\Psi)$
of the stress-energy tensor of the matter field appear in the r.h.s..
Therefore as the matter field $\Psi$ is discontinuous at the hypersurface
$\Sigma$ and as the kinetic terms no longer appear in the field equation for
the dilaton, one concludes that the dilaton $\varphi$ is necessarily
$C^1$ on $\Sigma$ -recall that the Maxwell field $F$ is at most discontinuous
accross $\Sigma$. The jump in the second derivatives of $\varphi$
accross $\Sigma$ is thus given by 
\begin{equation}
      [g^{\mu \nu} {\partial}_{\mu \nu} \varphi ]\:=
        \:-\frac{q \kappa k_F}{2} e^{-2 \kappa {\varphi}_{|\Sigma}} \lbrack F^2 \rbrack
        +2 q \kappa e^{2 \kappa {\varphi}_{|\Sigma} }[V(\Psi)]\hspace{1cm}.
\end{equation}

   It follows from this rapid investigation that the scalar fields in
scalar-tensor theories behave differently than the dilaton accross
a singular hypersurface: the former
are only $C^0$ while the latter is $C^1$ accross a singular hypersurface.
This is a consequence of the difference of their coupling to
the matter field. As the dilaton is $C^1$ it will not
contribute to the singular $\delta$-term appearing in the field
equations, and the expression for the surface stress-energy tensor  
of an arbitrary shell is not affected by the presence of the
dilaton and remains the same as in general relativity.\\
 
{\centerline{\sect{ACKNOWLEDGMENTS}}}
   For helpful discussions and suggestions we would like
to thank P. Forgacs, P.A. Hogan, W. Israel and B. Linet.
\\
\setcounter{equation}{0}
\renewcommand{\theequation}{A.\arabic{equation}}
\nappendix{\centerline{\sect{APPENDIX : STATIC SPHERICALLY SYMMETRIC SOLUTIONS}}}

	We present in this appendix some static spherically symmetric 
solutions of the Brans-Dicke theory. As they have an analog in general relativity
we call them by the same name.
\\
\centerline{{\it Analog of the Schwarzschild solution} }

As this solution has already been elsewhere -see Damour and  Esposito-Farese 
\cite{DEFB}- we only recall here
its main properties. In the Einstein-frame,
an exterior spherically symmetric solution satisfying the system of equations (3) and
(4) in vacuum ( $\psi\,=\,0$ ) is given by :
\begin{equation}
f(R)\,=\,{\left(1-\frac{a}{R}\right)}^{\frac{b}{a}} 
\end{equation}
\begin{equation}
	r^2 (R)\,=\,R^2\,\left(1-\frac{a}{R}\right)^{1-\frac{b}{a}}
\end{equation}
\begin{equation}
e^{2\varphi}\,=\,{\left(1-\frac{a}{R}\right)}^{\frac{d}{a}}\hspace{1cm},
\end{equation}
where $\,a\,$, $\,b\,$ et $\,d\,$ are positive constants, the values of which
are restricted by the condition
\begin{equation}
			a^2\,=\,b^2+d^2	\hspace{1cm}.
\end{equation}

	The above form of the metric is valid in the domain $\,R\,>\,a\,$.
This solution is the analog of the Schwarzschild metric with b playing
the role of the mass parameter.
We obtain an anolog of the Minkovski spacetime by putting 
$\,b=0\,$ in the above solution. Note that the resulting metric is
not flat because of the presence of the scalar field.
\\
\centerline{{\it Analog of the Reissner-Nordst\"{o}m solution}}
	In general relativity the Harisson transformation allows to
generate a charged solution of the Einstein-Maxwell fields equations
from a static vacuum solution. In the
same way, starting from the static spherically symmetric
uncharged solution obtained above (A1-4), one gets, using a similar transformation 
a family of static spherically symmetric charged solutions
indexed by an arbitrary non-zero parameter $\,\lambda\,$ and satisfying
the Brans-Dicke-Maxwell equations, which is given by
\begin{equation}
f(R)\,=\,g^{-2}(R)\,{\left(1-\frac{a}{R}\right)}^{\frac{b}{a}}\: ; \: \psi\,=\,0
\end{equation}
\begin{equation}
r^2\,(R)\,=\,g^2 \,(R)\,R^2\left(1-\frac{a}{R}\right)^{1-\frac{b}{a}}
\end{equation}
\begin{equation}
e^{2\varphi}\,=\,{\left(1-\frac{a}{R}\right)}^{\frac{d}{a}}\hspace{0.5cm},
\end{equation}
where the function g is defined by
\begin{equation}
	g(R)\,=\,{\cosh}^2 \lambda-{\left(1-\frac{a}{R}\right)}^{\frac{b}{a}}\,{\sinh}^2 \lambda\hspace{1cm}.
\end{equation}
The only non-zero component of the electromagnetic potential $A_\mu$ is equal to
\begin{equation}
	A_t \,=\,\frac{\left[\,{\left(1-\frac{a}{R}\right)}^{\frac{b}{a}}-1\,\right]
	\,\sinh(2\lambda)}{2\,g(R)}	
\end{equation}
which yields an electromagnetic field $F_{\mu \nu}$ with the only non-zero component
\begin{equation}
		F_{rt}\,=\,\frac{Q}{r^2}\hspace{1cm},		
\end{equation}
where $\,Q\,$ is the electric charge which is related to the parameter $\lambda$ as
\begin{equation}
		Q\,=\,\frac{b}{2}\,\sinh(2\lambda)\hspace{1cm}.
\end{equation}
It can be checked that putting $\lambda = 0$ in these
results gives back the analog of the Schwarzschild solution (A.1-4).
\newpage
  
\end{document}